\documentclass[twocolumn]{aastex631}

\usepackage{natbib}

\usepackage{amsthm}
\usepackage{graphicx}
\usepackage[space]{grffile}
\usepackage{latexsym}
\usepackage{amsfonts,amsmath,amssymb}
\usepackage{url}
\usepackage[utf8]{inputenc}
\usepackage{fancyref}
\usepackage{hyperref}
\hypersetup{colorlinks=true,citecolor=blue,urlcolor=blue,pdfborder={0 0 0},}
\usepackage{textcomp}
\usepackage{multirow,booktabs}
\usepackage{tablefootnote}
\usepackage{xspace}

\usepackage{xcolor}

\shorttitle{HST and VLT Reveal TRAPPIST-1 Frequent Microflares}
\shortauthors{Berardo}

\begin{document}
\title{Hubble's Multi-Year Search for Exospheres in the TRAPPIST-1 System Reveals Frequent Microflares}

\author[0000-0001-6298-412X]{David Berardo}
\affiliation{Department of Earth, Atmospheric and Planetary Sciences, Massachusetts Institute of Technology, Cambridge, MA 02139, USA}

\author[0000-0003-2415-2191]{Julien de Wit}
\affiliation{Department of Earth, Atmospheric and Planetary Sciences, Massachusetts Institute of Technology, Cambridge, MA 02139, USA}

\author[0000-0003-1462-7739]{Michael Gillon}
\affiliation{Astrobiology Research Unit, Université de Liège, Allée du 6 août 19, Liège, 4000, Belgium}

\author[0000-0002-0583-0949]{Ward S. Howard}
\affil{Department of Astrophysical and Planetary Sciences, University of Colorado, 2000 Colorado Avenue, Boulder, CO 80309, USA}
\affil{NASA Hubble Fellowship Program Sagan Fellow}

\author[0000-0002-9148-034X]{Vincent Bourrier}
\affiliation{Observatoire de Genève, Université de Genève, Chemin Pegasi 51, 1290 Versoix, Switzerland}

\author[0000-0003-4078-9328]{Matthew W. Cotton}
\affiliation{Department of Earth, Atmospheric and Planetary Sciences, Massachusetts Institute of Technology, Cambridge, MA 02139, USA}
\affiliation{Yusuf Hamied Department of Chemistry, University of Cambridge, CB2
1EW, United Kingdom}

\author[0000-0002-6236-2504]{Florian Quatresooz}
\affiliation{Department of Earth, Atmospheric and Planetary Sciences, Massachusetts Institute of Technology, Cambridge, MA 02139, USA}
\affiliation{ICTEAM Institute, Université Catholique de Louvain (UCLouvain), Louvain-la-Neuve, Belgium}

\author[0009-0007-1634-2792]{Léonie Hoerner}
\affiliation{Observatoire de Genève, Université de Genève, Chemin Pegasi 51, 1290 Versoix, Switzerland}

\author[0000-0001-5657-4503]{Emeline Bolmont}
\affiliation{Observatoire de Genève, Université de Genève, Chemin Pegasi 51, 1290 Versoix, Switzerland}
\affiliation{Centre sur la Vie dans l’Univers, Université de Genève, 1211 Geneva, Switzerland}

\author[0000-0003-2415-2191]{Artem Burdanov}
\affiliation{Department of Earth, Atmospheric and Planetary Sciences, Massachusetts Institute of Technology, Cambridge, MA 02139, USA}

\author[0000-0002-6523-9536]{Adam J.\ Burgasser}
\affiliation{Department of Astronomy \& Astrophysics, UC San Diego, La Jolla, CA, USA}

\author[0000-0002-9355-5165
]{Brice-Olivier Demory}
\affiliation{Center for Space and Habitability, University of Bern, Gesellschaftsstrasse 6, 3012 Bern, Switzerland}
\affiliation{Division of Space Research and Planetary Sciences, Physics Institute, University of Bern, CH-3012 Bern, Switzerland}
\affiliation{ARTORG Center for Biomedical Engineering Research, University of Bern, Murtenstrasse 50, CH-3008, Bern, Switzerland}

\author[0000-0001-9704-5405]{David Enhrenreich}
\affiliation{Observatoire de Genève, Université de Genève, Chemin Pegasi 51, 1290 Versoix, Switzerland}
\affiliation{Centre sur la Vie dans l’Univers, Université de Genève, 1211 Geneva, Switzerland}

\author[0000-0003-2805-8653]{Susan M.  Lederer}
\affiliation{NASA Johnson Space Center, 2101 NASA Parkway, Houston, Texas,
77058, USA.}

\author[0000-0002-3627-1676]{Benjamin V. Rackham}
\affiliation{Department of Earth, Atmospheric and Planetary Sciences, Massachusetts Institute of Technology, Cambridge, MA 02139, USA}
\affiliation{Department of Physics and Kavli Institute for Astrophysics and Space Research, MIT, Cambridge, MA 02139, USA}

\author[0000-0002-6892-6948]{Sara Seager}
\affiliation{Department of Earth, Atmospheric and Planetary Sciences, Massachusetts Institute of Technology, Cambridge, MA 02139, USA}
\affiliation{Department of Physics and Kavli Institute for Astrophysics and Space Research, MIT, Cambridge, MA 02139, USA}
\affiliation{Department of Aeronautics and Astronautics, MIT, 77 Massachusetts Avenue, Cambridge, MA 02139, USA}

\author[0000-0002-5510-8751]{Amaury Triaud}
\affiliation{School of Physics and Astronomy, University of Birmingham, Edgbaston, Birmingham, B15 2TT, United Kingdom.}

\correspondingauthor{David Berardo, Julien de Wit}
\email{berardo@mit.edu, jdewit@mit.edu}

\begin{abstract}
Ly-$\alpha$ observations provide a powerful probe of stellar activity and atmospheric escape in exoplanetary systems. We present here an analysis of 104 HST/STIS orbits monitoring the TRAPPIST-1 system between 2017 and 2022, covering 3--5 transits for each of its seven planets. 
We rule out transit depths $\gtrsim20\%$, which translates into an upper limit on the escape rate of $1064~EO_H$/Gyr for planet b ($1~EO_H$ is the Earth-ocean-equivalent hydrogen content), in agreement with recent claims that planet b should be airless. These upper limits are $\sim$3 times larger than expected from the photon noise due to a large baseline scatter, which we ultimately link to TRAPPIST-1's intrinsic Ly-$\alpha$ variability from frequent ``microflares.'' While JWST observations of TRAPPIST-1 in the near infrared have shown that $\sim10^{30}$-erg flares occur every $\sim$6 hours, we report here $\sim10^{29}$-erg flares on sub-hour timescales in the HST/STIS and also Very Large Telescope (VLT) $g^{'}$ observations. 
The FUV and optical amplitudes ($\sim$400$\%$ vs $\sim$3$\%$, respectively) for flares with similar waiting-times indicate flare temperatures of 11000$^{+4200}_{-3100}$~K over 0.011$^{+0.03}_{-0.01}$\% of the stellar disk. Finally, our multi-year baseline reveals a variability with $P = 3.27 \pm 0.04$ days, providing further validation of the previously reported 3.295-day rotation period for TRAPPIST-1. 
These results highlight the importance of accounting for stellar microvariability when searching for exospheres around active M dwarfs.

\end{abstract}

\keywords{}

\section{Introduction}

\begin{figure*}
	\centering
	\includegraphics[width=7in,angle=0]{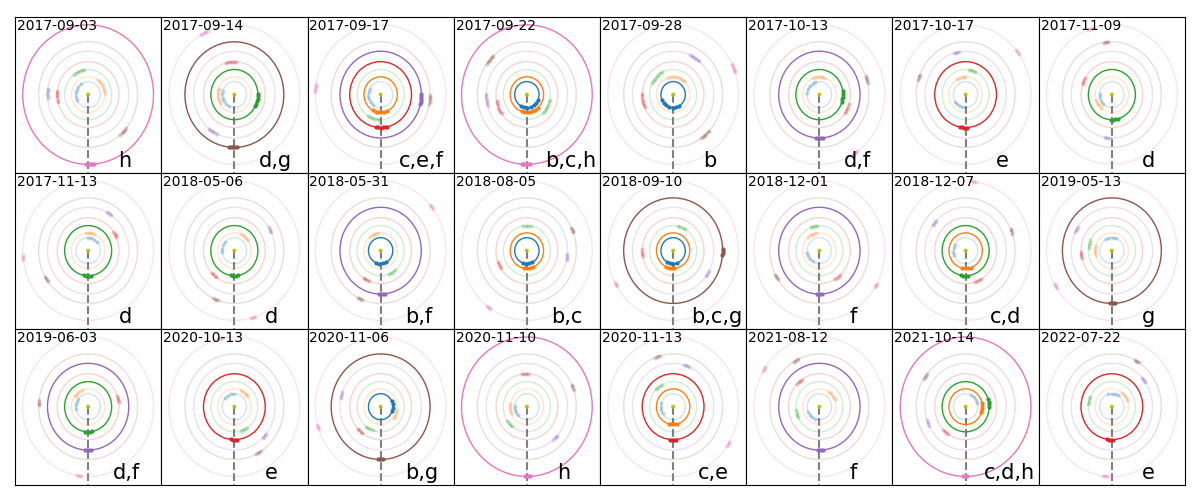}
	\caption{Overview of the orbital positions of the TRAPPIST-1 planets during each of the 24 visits. Planets which transit during a visit are in bold, with the observer line of sight represented by the dashed black line. The bottom right corner highlights which planet each orbit corresponds to.}
	\label{fig: observations}
\end{figure*}


The Hubble Space Telescope (HST) has played a pioneering role in the atmospheric exploration of terrestrial exoplanets. The favorable brightness and planet-to-star size ratios of the TRAPPIST-1 system \citep{gillon:2017} allowed for the search for H$_2$-dominated atmospheres around its seven planets using HST/WFC3 \citep{dewit:2016,dewit:2018,wakeford:2019,garcia:2022}. These observations ruled out the presence of extended, hydrogen-dominated atmospheres for all planets in the system, a result independently supported by mass--radius measurements of the planets \citep{turbet:2020,agol:2021}. 

The next frontier in the atmospheric exploration of the TRAPPIST-1 system involves searching for thinner, secondary atmospheres. Such atmospheres could be notably supported by large volatile reservoirs. Fortunately, TRAPPIST-1's planetary densities are consistent with scenarios involving iron depletion and/or volatile enhancement relative to Earth \citep[e.g.,][]{agol:2021,lichtenberg:2022}. Although the detection of such secondary atmospheres is currently only within reach with JWST \citep[e.g.,][]{kaltenegger:2009,dewit:2013,morley:2017,lustig-yaeger:2019,TJCI2023}---and remains challenged by stellar contamination
\citep{Lim2023, Radica2025}, HST retains an important and complementary role to play. Indeed, HST's unparalleled ultraviolet and visible-wavelength capabilities allow unique access to diagnostics of stellar activity and atmospheric loss, especially for close-in planets.

Planets that orbit sufficiently close to their host stars are at risk of significant atmospheric escape driven by intense stellar irradiation \citep{burrows:1995,guillot:1996,lammer:2003, yelle:2004, garciamunoz:2007, murray-clay:2009, owen:2012}. This process can produce an extended envelope or tail of hydrogen-dominated material that escapes, commonly known as an ``exosphere.'' 
Unlike atmospheres, exospheres are primarily detected via the narrow absorption features, such as the hydrogen Lyman-$\alpha$ (Ly-$\alpha$) line \citep{vidal-madjar:2004} or the meta-stable helium 2$^3$S absorption line at 1,083 nm \cite{spake:2018}. 

Previous observations have provided a preliminary assessment of the Ly-$\alpha$ emission from TRAPPIST-1. \cite{bourrier:2017} found that the stellar Ly-$\alpha$ emission evolved over a three-month span and used these data to assess the potential history of water loss in the system due to stellar irradiation. Their analysis suggested that planets b through d may still be undergoing runaway atmospheric escape, possibly having lost the equivalent of over 20 Earth oceans. In contrast, planets e through h may have lost significantly less---closer to 3 Earth oceans. Additional reconnaissance detected marginal flux decreases during transits of TRAPPIST-1b and c \citep{bourrier:2017a}, hinting at the possibility of extended hydrogen exospheres emanating from these two planets.

In this paper, we present the results of an extensive follow-up campaign of the TRAPPIST-1 system with 104 new orbits of HST/STIS observations conducted between 2017 and 2022 (HST-GO-15304, PI: de Wit). In Section 2, we introduce the data, detail our reduction and analysis, and report the derived constraints on the Ly-$\alpha$ transit depths and their interpretation. In Section 3, we investigate the origin of the unexpectedly large scatter in TRAPPIST-1's Ly-$\alpha$ flux, which we ultimately attribute to frequent (sub-hour) microflares. We discuss the implications of these findings and summarize our conclusions in Section 4.

\begin{figure*}
\centering

\includegraphics[width=7.5
    in,angle=0]{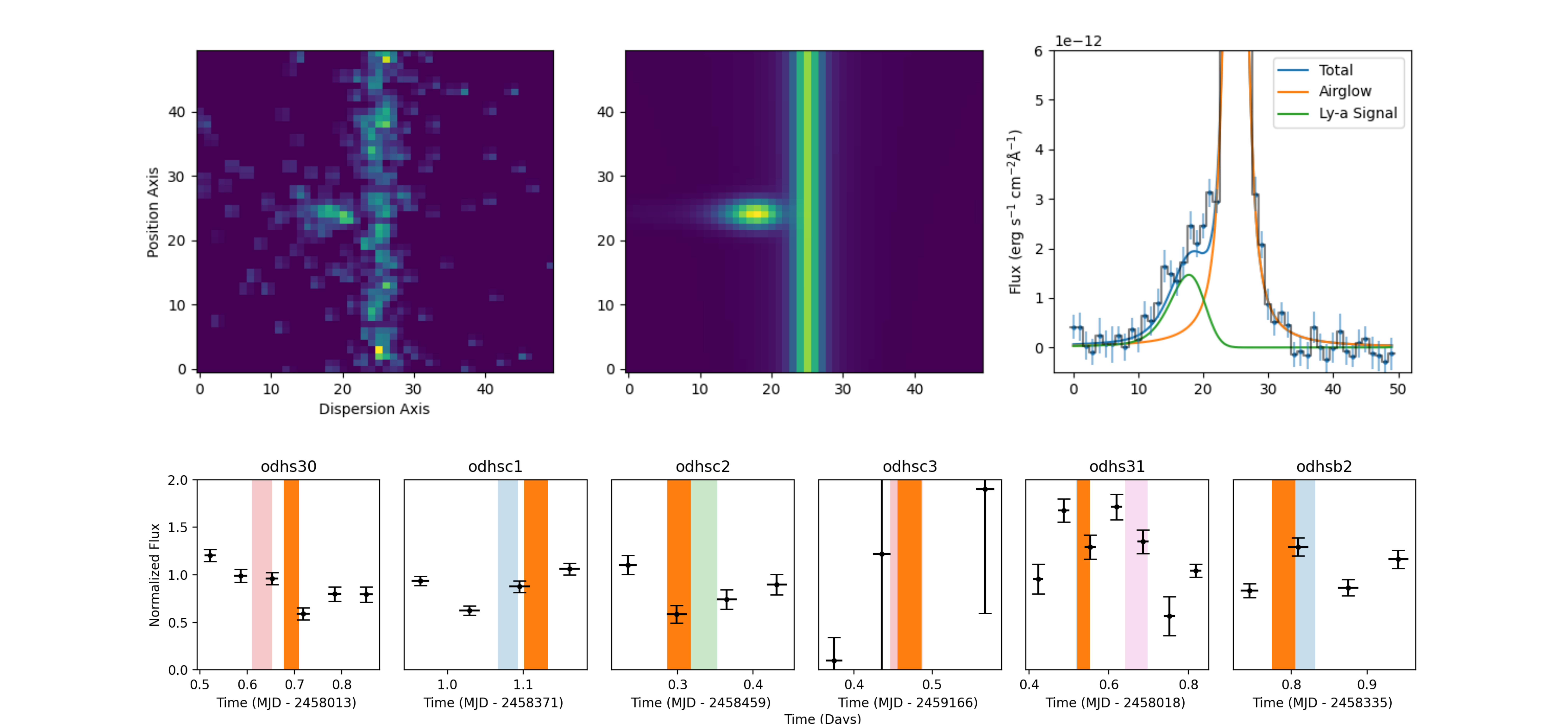}
\caption{Analysis \& Data Overview. (a) \textit{Top Left:}  Systematic-corrected signal image, for which a model fit is retrieved, which is shown in the central panel. \textit{Top Right:} Spatially-flattened spectral profile of the image, indicating the strong airglow peak and blue-wing of TRAPPIST-1's Ly-$\alpha$ emission, along with the best fit model in orange. Note that this is not the intrinsic Ly-$\alpha$ signal from TRAPPIST-1, but the signal after the effects of ISM absorption and STIS systematics.  (b) Ly-$\alpha$ time series for all visits containing a transit of TRAPPIST-1 c, shown in orange indicating the duration of its transit, normalized at the visit level. Other planets which happened to be transiting during these windows are also shown. In each panel the transit duration is the same, but the baseline of the visit is not constant leading to the change in width of the orange bar. All the time series are shown in \autoref{fig:all_transits}.}
\label{fig: transit fit example}
\end{figure*}


\section{Searching for Exospheres}

In this Section, we introduce the extensive HST/STIS dataset gathered from 2017 to 2022 in order to search for evidence of exospheres around each of the seven TRAPPIST-1 planets (HST-GO-15304, PI: de Wit). We first present the dataset and reduction methodology, then detail the modeling and PSF-fitting framework used to extract the Ly-$\alpha$ light curves. Finally, we present the results of our transit search and the corresponding constraints on planetary exospheres.

\subsection{Data}

The observations were taken with the STIS instrument using the G140M grating, which spans 1194--1249\,\AA\ with a resolution of $\sim$20\,km\,s$^{-1}$, providing full coverage of the 1216\AA\ Ly-$\alpha$ emission line. The observations were conducted from September 2017 to July 2022, with each visit coinciding with a transit of one or more planets (see \autoref{tab: STIS visits}) following the ``two-birds-one-stone'' strategy (as in HST-GO-14500 and 14873, PI: de Wit) to maximize the observation efficiency. 

Each visit included two to seven orbits of HST, with typically at least one orbit before and after the transit(s). Each orbit generally consisted of a single $\sim$2000-second integration.  \autoref{tab: STIS visits} summarizes the visits, including the number of orbits and transiting planets. In total, the dataset includes 3 transits of planets g, 4 of e and h, 5 of b, d, and f and 6 of c.


\autoref{fig: observations} provides a visual summary of the system geometry during each visit, highlighting which planets transited the star at that time. The left panel of \autoref{fig: transit fit example} presents a representative exposure of an TRAPPIST-1 observation with HST/STIS. The dominant feature in the image is the geocoronal hydrogen emission from the Earth's atmosphere \citep{meier:1970}, along with the weaker signal from the blue wing of TRAPPIST-1's Ly-$\alpha$ emission to its left (right panel).

\subsection{Model}

We model each observation as the sum of several components: a background signal, geocoronal airglow from Earth's atmosphere, and Ly-$\alpha$ emission from TRAPPIST-1, which has been attenuated by absorption from the interstellar-medium (ISM) \citep{landsman:1993,wood:2005}. We describe in \autoref{appendix:model} the mathematical formulation of each of these components, and how they are used to ultimately retrieve a Ly-$\alpha$ lightcurve of TRAPPIST-1. Given the low signal-to-noise ratio (SNR) of the TRAPPIST-1 emission and the proximity of the bright geocoronal airglow, we adopt point spread function (PSF) fitting---rather than aperture photometry---as our extraction method. This approach allows for more robust separation and modeling of overlapping spectral components.


\vspace{-00mm}
\subsection{PSF-Fitting Framework}

After an initial calibration of each visit (including frame rotation and outlier clipping), we used the \texttt{calstis} module\footnote{https://stistools.readthedocs.io/en/latest/calstis.html} to produce $mod_{x2d}$ files for model fitting. This processing step performs standard reduction tasks, such as flat-fielding, dark subtraction, cosmic ray removal, and wavelength calibration. 

To constrain the dominant airglow signal, we first performed an independent fit across all files. The airglow SNR is orders of magnitude higher than the TRAPPIST-1 emission and originates mostly from spatial rows unaffected by the stellar signal. 
Once we derived tight priors on the airglow's position and shape, we conducted an initial first round of Markov chain Monte Carlo \citep[MCMC;][]{metropolis:1953,hastings:1970} fitting using the full model, including contributions from the background, airglow, and Ly-$\alpha$ line. 

We then used the output of this fit to create an error map and applied targeted $>4\sigma$ clipping to remove local outliers, including those associated with TRAPPIST-1's signal. A final run of MCMC fitting provided posterior distributions for all model components, including TRAPPIST-1's Ly-$\alpha$ flux.

\autoref{fig: transit fit example} shows a representative fit, illustrating the combined signal of background, airglow, and TRAPPIST-1's Ly-$\alpha$ line. The bright vertical band originates from geocoronal airglow across all spatial positions, while the smaller, fainter feature is from TRAPPIST-1 (in particular the blue wing of its Ly-$\alpha$ line).



\subsection{No Detection of Transits in Ly-$\alpha$}

\autoref{fig: transit fit example} show the Ly-$\alpha$ light curves for all visits that included a transit of planet c. Each panel corresponds to a different visit, with the transit window of planet c highlighted. Since some visits have more orbits than others, the apparent duration of the transit is c is smaller, but this is just due to an increased observing baseline. Transits of other planets during the same visits are also indicated. Additional light curves for the remaining planets are shown in \autoref{APP: lightcurves}.

For each visit, we defined the transit depth as the relative difference between the in-transit and out-of-transit (OOT) flux. The uncertainty on the Ly-$\alpha$ transit depth was estimated by adding the standard deviations of the in-transit and OOT fluxes in quadrature. We applied this procedure to each visit where a given planet transited and then computed the average depth and associated uncertainty across all visits. The results are summarized in \autoref{tab: ly a transit depths}.

None of the planets show a statistically significant Ly-$\alpha$ transit signal (\autoref{fig: transit fit example} and \autoref{APP: lightcurves}). 
In addition to the transit non-detections, our results on this large dataset reveal a substantial variability in the data that does not correspond to the planet transit timings. This excess scatter---typically 3 to 5 times larger than the expected photon noise---is the bottleneck for our search for exospheres in the TRAPPIST-1 system. We investigate its origin in Section 4.


\subsection{Limits on the Presence of Exospheres around TRAPPIST-1's Planets}

Given the unexpectedly high scatter of TRAPPIST-1's Ly-$\alpha$ flux, our sensitivity is limited to detecting transits with depths $\gtrsim$20\% at the $5\sigma$ level (\autoref{tab: ly a transit depths}). The corresponding $2\sigma$ upper limit of ${\sim}30\%$ for all planets is comparable to the constraint achieved for K2-25\,b \citep{rockcliffe:2021}. While non-detections can provide tight upper limits in quieter systems \citep[e.g., ${\sim}3\%$ at $2\sigma$ for HD~63433\,b,][]{zhang:2022}, the detectability of exospheres with HST/STIS varies greatly across targets, and so do the insights that can be drawn from non-detections. 

In the case of TRAPPIST-1, the possibility remains that shallow transits from extended hydrogen exospheres occur at depths of a few percent. However, such signals lie below the current sensitivity threshold of HST.
In other systems, non-detections have been explained by the efficient ionization of escaping hydrogen, which suppresses Ly-$\alpha$ absorption \citep[e.g., HD~97658~b,][]{bourrier:2017}. For the TRAPPIST-1 planets, the absence of large (5--8$\times R_p$) exospheres places only modest constraints and may similarly reflect ionization-driven suppression of the neutral hydrogen signature. 


We estimated the atmospheric mass loss from the TRAPPIST-1 planets following the method from \cite{bolmont:2017,bourrier:2017b,ribas:2016}, using Eq.~1 of \citet{bourrier:2017b} with an efficiency factor $\epsilon$ from \citet[][Fig.~2]{bolmont:2017} that depends on the XUV flux the planet receives. 
From the measured Ly-$\alpha$ luminosity of $7.93\times10^{27}$~erg.s$^{-1}$ (see Section~\ref{subsec:microflares}), we estimated an XUV luminosity of $1.83\times10^{28}$~erg.s$^{-1}$ using the relation of \cite{linsky:2014}.

Compared to earlier estimates from \cite{bolmont:2017,bourrier:2017b}, our updated mass-loss estimates are about 100-450 times larger. 
This is driven by two differences: (1) the inferred XUV luminosity in this study is $\sim 30$ times higher than the value used in 2017  ($6.28\times10^{26}$~erg.s$^{-1}$), which increases the mass-loss rate by a factor of 7--15 due to the decrease of $\epsilon$ for high XUV fluxes; and (2) we account here for a potentially high XUV cross-section given by $(R_{\rm XUV}/R_\star)^2 = 0.10$, whereas previous studies took it to be equal to the planetary radius. 
The second point leads to an increase in mass loss by an additional factor of 13--30 (larger for smaller planets).

These assumptions lead to water loss rates (as defined by Eq.~8 of \citealt{bolmont:2017}), expressed in Earth-ocean-equivalent hydrogen contents ($EO_H$), as high as $533~EO_H/$Gyr for planet b and exceeding $100~EO_H$/Gyr for all planets. 
The enhanced XUV cross-section has an especially pronounced effect on the smaller planets; for example, planet d may lose more than $650~EO_H$/Gyr under this assumption.
These values are globally doubled if we consider a XUV cross-section corresponding to $(R_{\rm XUV}/R_\star)^2 = 0.20$ (leading to a loss of $1064~EO_H$/Gyr for planet b). 

We note that the adopted value of $(R_{\rm XUV}/R_\star)^2 = 0.10$ represents an upper limit. If the true cross-section is smaller, the actual water loss could be significantly reduced. 
This parameter is particularly important as it influences the mass flux (via the $(R_{XUV}/R_p)^2$ factor of Eq.~1 of \citealt{bourrier:2017b}). 
However, it was also argued that taking $R_{\rm XUV} > R_{\rm p}$ leads to an overestimation of loss \citep{selsis:2007}.
Additionally, such high escape rates should be limited by the availability of water on the planet. 
Once a potentially high initial amount of water is lost \citep[e.g.][]{unterborn:2018}, the loss would be limited by the degassing rate of water.
Hence, as with previous studies \citep{ribas:2016,bolmont:2017,bourrier:2017b}, our estimates should be considered upper limits on water loss, and more precise escape simulations will be necessary to refine these values.

\section{TRAPPIST-1's Ly-$\alpha$ Variability: a Sign of Frequent Micro-Flares}

Considering that this extensive search for exospheres in the TRAPPIST-1 system is substantially limited by the apparent variability of TRAPPIST-1's Ly-$\alpha$ line, in this section, we investigate the possible origins for such variability.

\begin{figure*}[bt!]
	\hspace{-05mm}\includegraphics[width=7.2
    in,angle=0]{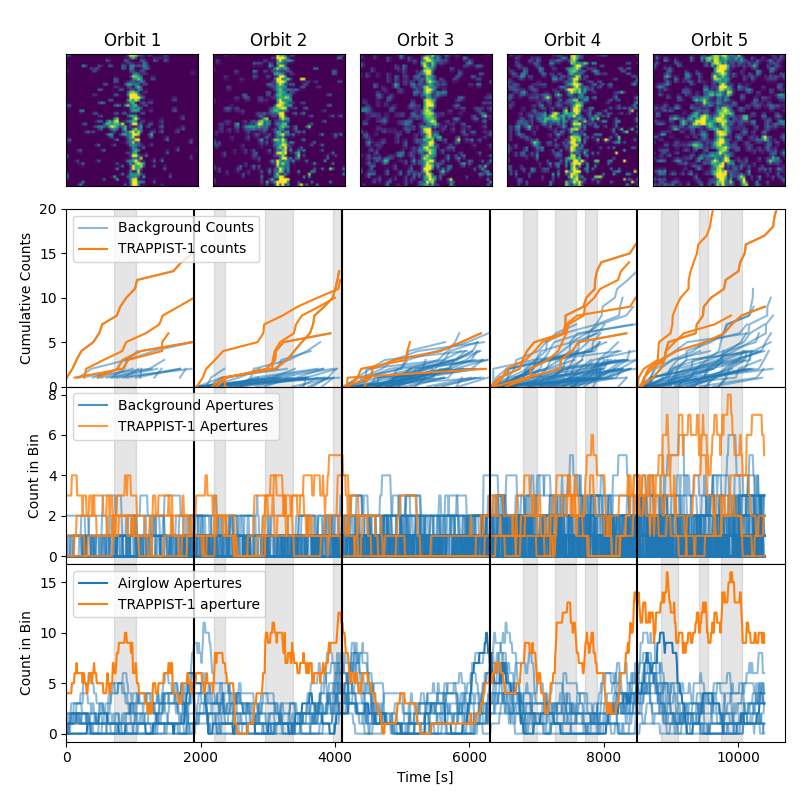}
	\caption[]{Signs of frequent flares in TRAPPIST-1 HST/STIS TIME-TAG data. Count rates for individual sections on the detector based on photon TIME-TAG data for visit odhs33. Solid black lines denote time between orbits. The top row of orbit-integrated exposures shows a 50x50 pixel stamp centered on TRAPPIST-1, highlighting the variability of the Ly-$\alpha$ emission. The vertical line of emission to the right of the signal is the airglow emission from the Earth. Top Panel: Cumulative photon counts for 4 sub-apertures centered around on TRAPPIST-1's Ly-$\alpha$ blue wing (orange curves) and detector background (not including airglow or Ly-$\alpha$ signal, blue curves). Middle panel: Count rate in a 8-pixel/5 minute window, shown for the same Ly-$\alpha$ apertures in orange and background apertures in blue. Bottom Panel: Similar count rates, but this time for a single aperture focused on the Ly-$\alpha$ of TRAPPIST-1 (orange curves) compared to the airglow signal (blue curves). Pale grey regions highlight time periods where the Ly-$\alpha$ shows an increased count rate above the background or airglow rates. Orbit 3 presents no signs of flares and the associated exposure do not show TRAPPIST-1's Ly-$\alpha$ signal suggesting that the bulk of it may be due to such sub-hour flares. We derive an upper limit on TRAPPIST-1's quiescent Ly-$\alpha$'s blue wing of 3e-12{$\mathrm{erg\ s^{-1}\ cm^{-2} \AA^{-1}}$} (see also \autoref{fig: la emission folded}).}
    \label{fig: time tagged}
\end{figure*}

\subsection{Intrinsic Versus Extrinsic Sources of Variability}

We considered several possible origins for the elevated intra-visit scatter seen in TRAPPIST-1's Ly-$\alpha$ lightcurves. One possibility is contamination from variable airglow or background noise. To test this, we examined the best-fit values for the Ly-$\alpha$ signal amplitude, the background parameters, and the airglow amplitude for each orbit within a visit and computed the correlation matrix among them. We find that the Ly-$\alpha$ amplitude has a typical correlation coefficient of $<0.5$  with both the background and airglow signals.

The lack of strong correlations with external parameters points to TRAPPIST-1 itself as the origin of the observed Ly-$\alpha$ variability. This suggests that the stellar emission varies on timescales comparable to an HST orbit. A possible explanation for the variation is frequent low-energy flaring activity occurring on sub-hour timescales, leading to orbit-to-orbit variations in the measured Ly-$\alpha$ flux.

\subsection{Time-Resolved Insights from STIS TIME-TAG Data}

To assess whether the excess scatter in TRAPPIST-1's Ly-$\alpha$ flux arises from intrinsic stellar activity, we leveraged a special property of the present STIS data: its TIME-TAG nature. The TIME-TAG mode records the arrival time and location of each detected photon event, enabling high time resolution within each $\sim$45-min integration---sufficient to search for frequent flares. In the TIME-TAG data, flares would manifest as transient increases in count rate for the pixels within the target's Ly-$\alpha$ line. 

Given the faintness of TRAPPIST-1's Ly-$\alpha$ emission and the visit-dependent systematics (e.g., airglow contamination), we focused on the eight visits with the highest SNR. We spatially binned the data into grids of $5{\times}5$, $8{\times}8$, and $10{\times}10$ pixels and examined the time series of count rates for the four bins covering the star's Ly-$\alpha$ signal (specifically the blue wing). We compared these to the count-rate timeseries for regions dominated by the background and airglow signals (see Figure\,\ref{fig: time tagged}). Count rates were calculated as the number of photon counts per spatial bin in successive five-minute intervals (roughly the median time between counts in an $8{\times}8$ background pixel bin for these high-SNR visits). 

We identified candidate flares as epochs where the summed count rate from the four bins covering TRAPPIST-1's Ly-$\alpha$ signal significantly exceeded the baseline trend, provided there was no concurrent spike in the airglow count rate (which typically increases toward the end of each orbit). In this way, we used the high-SNR airglow as a reference for time-dependent systematics and flagged outliers in TRAPPIST-1's Ly-$\alpha$ count rate as flare candidates. 

We find a median of three such flare-like events per visit, with amplitudes between 300 and 500$\%$. This rate is consistent with the hypothesis that flares on a timescale of the order of an orbit (i.e., $\sim$45\,min) drive the excess orbit-to-orbit scatter in the Ly-$\alpha$ flux. If flare occurrence were significantly more or less frequent than the orbital timescale, the resulting scatter would be lower. Notably, this inferred cadence ($O(45min)$) is much higher than the $O(6hrs)$ frequency reported from NIR measurements with JWST, suggesting we are probing a more frequent and less energetic segment of the flare population that is more readily detected at ultraviolet wavelengths. To further test this interpretation, we turned to ground-based optical monitoring of TRAPPIST-1 to search for corresponding microflaring activity.

\subsection{Independent Detection of Frequent Micro-Flaring Activity with the VLT}

In 2018, four transit windows of TRAPPIST-1\,c were observed with the FORS2 instrument on the UT1-Antu telescope of the Very Large Telescope (VLT) facility at the European Southern Observatory (ESO) on Cerro Paranal, Chile (Program 0102.C-0635, PI: M. Gillon, Gillon et al., in prep). These observations were conducted using the $g_\textrm{HIGH}$ filter (spectral range: 388--548\,nm) and in the 200\,KHz, $2{\times}2$ read-out mode of the MIT detector of the FORS2 instrument (pixel scale: 0.25'', field of view: $6.8' {\times} 6.8'$). The first transit (15 May 2018) was observed with 80-second exposures, while the subsequent transits (13 and 30 Nov 2018, 21 Nov 2019) used 100-second exposures to optimize the signal-to-noise ratio and duty cycle. All observations were obtained under clear sky conditions. 

Standard image calibration was applied, and aperture photometry was used to measure the flux of TRAPPIST-1 and of several comparison stars. Differential photometry revealed frequent, flare-like variability with amplitudes ranging from 1 to 50\% in all light curves. These events occurred at a rate of 1--2 per hour, consistent with the flare frequency inferred from our HST/STIS analysis. 

\begin{figure}[ht!]
\vspace{-2mm}\hspace{-8mm}\includegraphics[trim={15.9cm 5mm 0 1cm},clip,width=1.15\columnwidth]{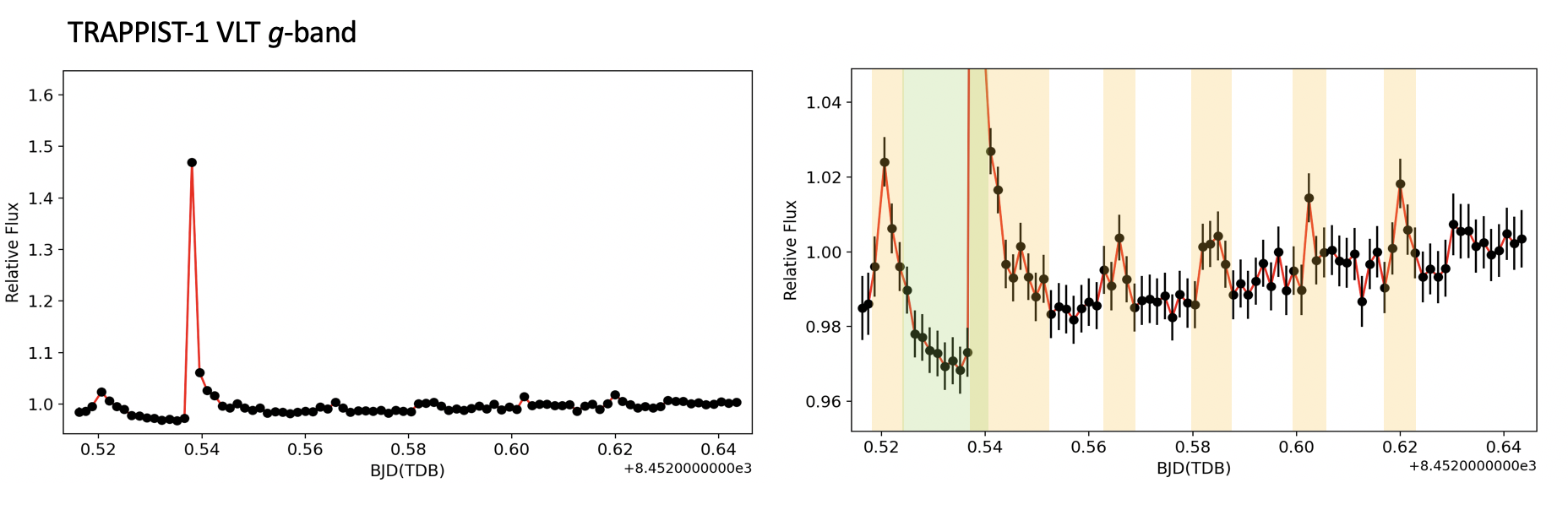}
\caption[VlT g-band photometry]{$g$-band light curve measured from the observations of TRAPPIST-1 by VLT on 30 Nov 2018. A flare of $\sim$50\% amplitude is visible (trimmed here). The orange bands show flare-like structures identified by eye in the light curve. The green band marks the expected transit window of TRAPPIST-1\,c. }
	\label{fig:vlt}
\end{figure}

As an illustration, \autoref{fig:vlt} shows the light curve of 30 Nov 2018. In addition to a transit of planet c and a gradual upward trend likely due to differential extinction, six flux increases lasting a few minutes are clearly visible.  

\subsection{Properties of TRAPPIST-1's Microflares}\label{subsec:microflares}
\begin{figure*}
	\centering
        \includegraphics[width=\textwidth]{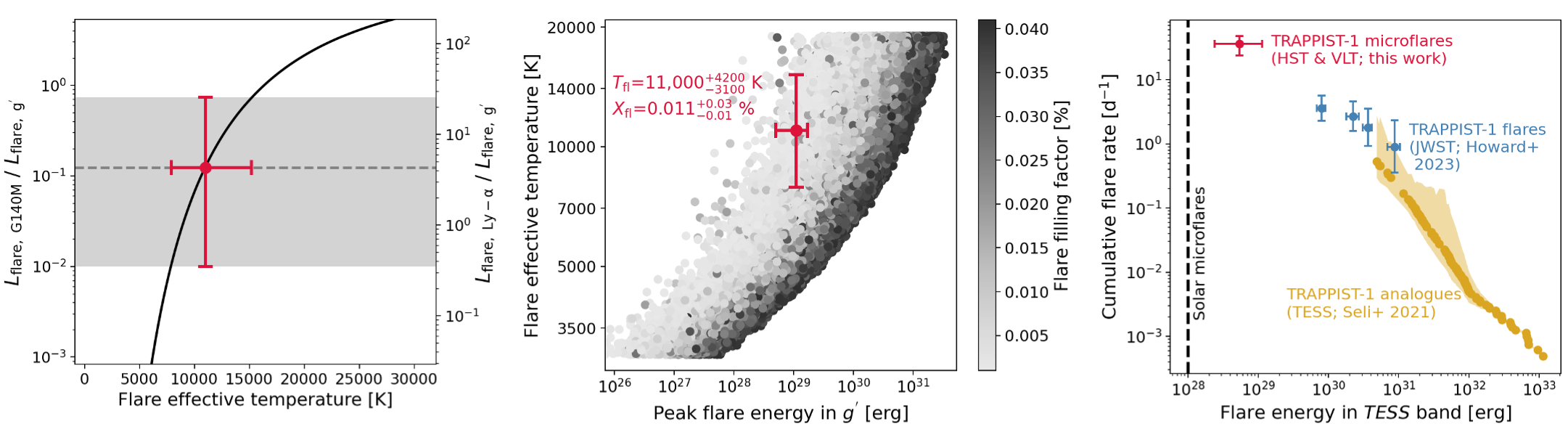}
	\caption[Microflare properties.]{Properties and occurrence rate of TRAPPIST-1 microflares. Left: The observed ratio of G140M to $g^{'}$ flare continuum and uncertainty (gray) compared against that of a blackbody of $T_\mathrm{fl}$ (black) gives $T_\mathrm{fl}$=11000$^{+4200}_{-3100}$~K. Middle: Position of our microflares relative to simulated flares across a range of energies, temperatures, and filling factors on the stellar disk. Right: Cumulative flare frequency distribution of TRAPPIST-1 and analogs, each shown in \textit{TESS} band. The estimated microflare \textit{TESS} energy and rate (red) reveal these events are a decade smaller and more frequent than flares regularly observed from TRAPPIST-1. This is analogous to how solar microflares are at least a decade smaller than typical C-class solar flares \citep{hannah:2011}.}
	\label{fig:microflare_panels}
\end{figure*}

The standard approach for comparing flare properties across different wavelength regimes in the absence of simultaneous data relies on the assumption that flares with similar waiting times share underlying emission mechanisms (e.g., \citealt{loyd:2018, jackman:2023, howard:2025}). Following this framework, we estimate the flare effective temperature ($T_\mathrm{fl}$) and filling factor ($X_\mathrm{fl}$) responsible for the observed peak flux enhancements of $\sim$400\% in Ly-$\alpha$ and 4\% in the $g^{'}$ band. Using the known stellar spectrum and distance, we convert these enhancements to luminosities of $L_\mathrm{Ly-\alpha}$= 7.93$\times$10$^{27}$ and $L_g'$=4.23$\times$10$^{26}$ erg s$^{-1}$. We multiply the luminosities by the Ly-$\alpha$ and optical peak durations of 60$\pm$20 and 260$\pm$130 s to obtain peak energies of $E_\mathrm{Ly-\alpha}$= 4.8$\pm$1.6$\times$10$^{29}$ erg and $E_g'$= 1.1$\pm$0.6$\times$10$^{29}$ erg, where the uncertainties depend on the peak duration. These small flares are at least an order of magnitude more energetic than typical solar microflares, which emit between 10$^{26}$--10$^{28}$ erg \citep{hannah:2011}. We nevertheless refer to them as ``microflares,'' given their high occurrence rate relative to classical flares on TRAPPIST-1 and the Sun \citep{Murray2022,howard:2023}. 

To estimate the flare continuum in the STIS G140M bandpass, we adopt the ratio of continuum to Ly-$\alpha$ emission of 0.03$^{+0.14}_{-0.02}$ from simultaneous line and continuum FUV flare observations \citep[][Table 8]{loyd:2018}, yielding $E_\mathrm{G140M}$=1.4$^{+6.7}_{-1.1}\times$10$^{28}$ erg. The uncertainty in $E_\mathrm{G140M}$ is dominated by the large uncertainty in the Ly-$\alpha$ to continuum scaling relation. Using  $E_\mathrm{G140M}$ and $E_{g^{'}}$, we solve for the one-component blackbody curve consistent with both energies, finding $T_\mathrm{fl}$=11000$^{+4200}_{-3100}$~K and $X_\mathrm{fl}$=0.011$^{+0.03}_{-0.01}$\%. Our constraints on the flare temperature and filling factor place the TRAPPIST-1 events within the typical ranges of 9000--14,000~K and 0.005--0.5\% seen for most solar/stellar flares \citep{kowalski:2013}. The inferred filling factor also aligned with values reported for TRAPPIST-2 flares in the 0.6--2.8~$\mu$m range from NIRISS observations, while the temperature is notable higher than the 3000--5000 K values derived for TRAPPIST-1 flares observed at red optical/IR wavelengths \citep{maas:2022, howard:2023}.




\begin{figure*}[bt!]
\hspace{-03mm}	\includegraphics[width=7.1in,angle=0]{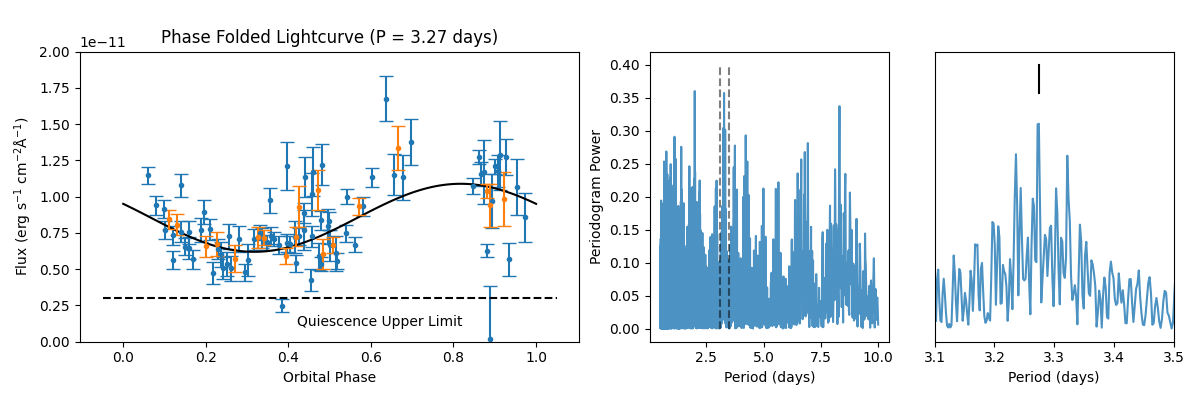}
	\caption[TRAPPIST-1 Long Term Flux Trends]{Complementary insights into TRAPPIST-1 from HST/STIS's multi-year monitoring. 
    Left: Absolute emission flux from TRAPPIST-1, folded to various periods. The blue data points are for each individual orbit, while the orange points are binned to show the mean flux value for each visit. The upper limit on TRAPPIST-1's quiescent Ly-$\alpha$'s blue wing of 3e-12{$\mathrm{erg\ s^{-1}\ cm^{-2} \AA^{-1}}$} is shown as an horizontal dashed line (see also \autoref{fig: time tagged}). Center: Lomb-Scargle periodogram power, centered on the current best estimate for the rotation period of TRAPPIST-1 of 3.295 days. Right: Zoomed in view centered around a period of 3.29 days \citep{vida:2017}.}
    \label{fig: la emission folded}
\end{figure*}

\section{Independent Constraint on TRAPPIST-1's Rotation Period}

We leveraged the long, multi-year baseline of our HST/STIS monitoring campaign to search for periodic signals in the out-of-transit Ly-$\alpha$ flux from TRAPPIST-1. We combined the absolute Ly-$\alpha$ flux measurements for each orbit and computed a Lomb-Scargle periodogram \citep{lomb:1976,scargle:1982}. Among several significant peaks, we identified a strong periodicity at   $P = 3.27 \pm 0.04$ days (\autoref{fig: la emission folded}), which closely matches the previously reported stellar  rotation period of 3.295 days from K2 photometry \citep{vida:2017}. Another important period found in this 5-yr long monitoring is Earth's year (365.25 days), clearly identified by periodicity in the peak position and amplitude of the airglow \autoref{fig:airglow variation}. 

To further assess the rotation signal, we also performed an independent MCMC analysis of the Ly-$\alpha$ time series. The results similarly exhibited a multi-modal distribution of period values, with a central value of 3.26 $\pm$ 0.03 days.  \autoref{fig: la emission folded} shows the stellar flux folded to the highest likelihood period of 3.274 days, which reveals coherent modulation despite substantial orbit-to-orbit scatter.

To test the robustness of the detected signal, we performed an injection--recovery test. We injected varying sinusoidal modulations with a range of amplitudes and periods, using the actual observing times and out-of transit scatter measured for each visit. In all cases, the injected rotation period was recovered as either the dominant or second most significant peak in the periodogram, confirming that the dataset is sensitive to periodicities on this timescale and supporting our detection of TRAPPIST-1's rotation period.

\section{Conclusions}


We have presented the results of a five-year HST/STIS campaign (104 orbits, 24 visits) aimed at detecting extended neutral hydrogen exospheres around the seven planets in the TRAPPIST-1 system via Ly-$\alpha$ transit observations.
This study represents one of the most extensive UV monitoring efforts of any M dwarf exoplanetary system to date.

Our analysis yielded no statistically significant Ly-$\alpha$ transit detections for any planet (\autoref{tab: ly a transit depths}).
We rule out transit depths $\gtrsim20\%$ at the 5$\sigma$ level and place $2\sigma$ upper limits of ${\sim}30\%$ on the presence of large exospheres---comparable to previous constraints for similar systems such as K2-25\,b \citep{rockcliffe:2021}.
These limits are substantially degraded relative to the photon noise floor, owing to an unexpectedly high level of variability in the out-of-transit stellar flux.

We identify this excess scatter as a manifestation of intrinsic stellar variability, consistent with frequent, short-duration microflares.
Using STIS TIME-TAG data, we detect sub-hour-timescale increases in the Ly-$\alpha$ count rate across multiple visits.
These events are independently corroborated by g-band observations with VLT/FORS2, which show 1--2 flares per hour with amplitudes up to 50\%.
The relative strengths of the Ly-$\alpha$ and visible signals suggest flare temperatures of $O(10^4)$\,K covering a few percent of the stellar surface.

These microflares not only limit the detectability of shallow transit signals in the UV, but also highlight the need for new strategies to characterize exospheres for planets around active M dwarfs.
In the case of TRAPPIST-1, while we cannot rule out small, escaping atmospheres (with ${\sim}$1--10\% transits), their detection is currently beyond the sensitivity of HST.

In addition to these constraints, our large dataset enables two further insights.
First, we construct a high-SNR composite STIS spectrum of TRAPPIST-1 from 1204--1232\,\AA{}, which reveals potential additional emission features.
Second, we recover a stellar rotation period of $3.27 \pm 0.04$\,d from the long-term Ly-$\alpha$ variability, consistent with previous measurements from optical monitoring \citep{vida:2017}.

In total, this study illustrates both the promise and limitations of Ly-$\alpha$ transit spectroscopy for detecting exospheres in M-dwarf systems. Non-detections set \textit{context-dependent} limits whereby null results must be interpreted relative to each system's unique noise profile. Namely, this study underscores the importance of accounting for stellar variability---especially previously underappreciated microflare activity---for UV observations, and it highlights the value of complementary observations (VLT/FORS2 g-band observations) for understanding stellar activity and thus probing exoplanetary mass loss effectively in the context of M dwarfs.

\section*{Acknowledgments}
This work is based on observations made with the NASA (National Aeronautics and Space Administration)/European Space Agency (ESA) Hubble Space Telescope (HST), obtained at the Space Telescope Science Institute (STScI), which is operated by the Association of Universities for Research in Astronomy, Incorporated, under NASA contract NAS5-26555. These observations are associated with program GO-15304 (PI: de Wit), support for which was provided by NASA through a grant from the STScI. D.B. and J.d.W. acknowledge support from the European Research Council (ERC) Synergy Grant under the European Union’s Horizon 2020 research and innovation program (grant No. 101118581 — project REVEAL). B.V.R. acknowledges additional support for program number HST-AR-17551 through a grant from the STScI under NASA contract NAS5-26555. W.H. received funding through the NASA Hubble Fellowship grant HST-HF2-51531 awarded by STScI, which is operated by the Association of Universities for Research in Astronomy, Inc., for NASA, under contract NAS5-26555. 
E.B. acknowledges the financial support of the SNSF (grant number: 200021\_197176 and 200020\_215760) and that this work was carried out within the framework of the NCCR PlanetS, supported by the Swiss National Science Foundation under grants 51NF40\_182901 and 51NF40\_205606.
Finally, the MIT crew acknowledges the unparalleled support of the Earth Resources Laboratory's new coffee machine.

\bibliography{ms}

\appendix{}
\section{List of Observations}

\begin{table*}[ht]
\centering
\begin{tabular}{lccccc}
\hline
\multicolumn{1}{l}{Visit Id} & \multicolumn{1}{c}{\# orbits} & \multicolumn{1}{c}{Start (MJD)}& \multicolumn{1}{c}{End (MJD)}& \multicolumn{1}{c}{Transits}& \multicolumn{1}{c}{Mean Visit Flux } \\ 
\multicolumn{1}{l}{} & \multicolumn{1}{c}{} & \multicolumn{1}{c}{(MJD)}& \multicolumn{1}{c}{(MJD)}& \multicolumn{1}{c}{}& \multicolumn{1}{c}{($\mathrm{erg\ s^{-1}\ cm^{-2} \AA^{-1}}$) } \\ 
\hline

odhs07 & 6 & 57999.80554628 & 58000.13234036 & h & 9.85e-12\\
odhs29 & 6 & 58010.53201628 & 58010.85977110 & g & 5.95e-12\\
odhs30 & 6 & 58013.51192369 & 58013.83968999 & c,e & 8.45e-12\\
odhs31 & 7 & 58018.41267591 & 58018.80671554 & b,c,h & 2.03e-11\\
odhs32 & 7 & 58024.30720220 & 58024.70121849 & b & 7.83e-12\\
odhs33 & 5 & 58039.00752627 & 58039.26906553 & f &  2.37e-11\\
odhs16 & 4 & 58043.97447072 & 58044.16982961 & e & 5.89e-12\\
odhs05 & 4 & 58066.49076783 & 58066.68606894 & d & 5.72e-12\\
odhs08 & 4 & 58070.46400857 & 58070.65934413 & d & 6.05e-12\\
odhsd1 & 4 & 58244.62177988 & 58244.81603010 & d & 1.50e-10\\
odhsb1 & 4 & 58269.25489373 & 58269.44886373 & b,f & 1.38e-10\\
odhsb2 & 4 & 58335.73444453 & 58335.92925934 & b,c & 6.41e-12\\
odhsc1 & 4 & 58371.95377446 & 58372.14935298 & b,c & 9.34e-12\\
odhsf1 & 4 & 58453.33295479 & 58453.52589479 & f & 9.31e-12\\
odhsc2 & 4 & 58459.22383442 & 58459.41937850 & c,d & 6.74e-12\\
odhsg1 & 4 & 58616.00951054 & 58616.20457980 & g & 1.74e-10\\
odhsf2 & 4 & 58637.46029758 & 58637.65537832 & d,f & 1.33e-11\\
odhse1 & 4 & 59135.84576521 & 59136.03995521 & e & 1.04e-11\\
odhsg2 & 4 & 59159.54331234 & 59159.73883345 & g & 8.05e-12\\
odhsh1 & 4 & 59163.38209716 & 59163.57863642 & h & <5.40e-12\\
odhsc3 & 4 & 59166.36554642 & 59166.55782642 & c,e & 4.77e-12\\
odhsf3 & 4 & 59438.47639171 & 59438.67289652 & f & 7.18e-12\\
odhsh2 & 4 & 59501.15662499 & 59501.35231943 & h & 1.0e-11\\
odhse3 & 4 & 59782.36577810 & 59782.56156514 & e & 7.17e-12\\
odhs33 & 5 & 58039.00752627 & 58039.26906553 & f &  1.55-11\\
odhs31 & 7 & 58018.41267591 & 58018.80671554 & b,c,h & 1.03e-11\\
odhsd1 & 4 & 58244.62177988 & 58244.81603010 & d & 1.50e-11\\
odhsb1 & 4 & 58269.25489373 & 58269.44886373 & b,f & 1.38e-11\\
odhsg1 & 4 & 58616.00951054 & 58616.20457980 & g & 3.74e-12\\
odhsh1 & 4 & 59163.38209716 & 59163.57863642 & h & 9.40e-12\\

\hline
\end{tabular}
\caption{Description of the observations taken as part of the observing program. Times listed are start of the first orbit and end of the last orbit with a visit, comprising anywhere from 4 to 7 individual orbits/integrations. The `Transits' column lists which of the TRAPPIST-1 planets transits during the visit (typically during the central orbit). The mean flux values are those used to normalize the lightcurves for each visit shown in figure \ref{fig:all_transits}}
\label{tab: STIS visits}
\end{table*}

\newpage

\section{Further Model Description}
\label{appendix:model}

\subsection{Airglow Model}
The dominant signal observed is from geocoronal hydrogen emitting Ly-$\alpha$ photons around the Earth, which extends well past the orbit of the moon \citep{baliukin:2019}. We model the airglow signal with a Voigt profile \citep{sim:2001} across the wavelength axis (the x-axis in \autoref{fig: transit fit example}), using the implementation found in the astropy package \citep{astropy:2013,astropy:2018,astropy:2022}. Along the y-axis (spatial) we assume the position of the airglow to be constant.

The Voigt profile is defined using a positional offset $x_{airglow}$, an amplitude $A_{airglow}$, and the full-width half-maximum (FWHM) of both the Gaussian and Lorentzian profiles, $\mathrm{fwhm}_{G}$ and $\mathrm{fwhm}_{L}$ respectively. These FWHM values are taken to be constant across all observations, and are thus fixed during fitting, having been measured during preliminary global fits to be 1.83458 for the 
Lorentzian width and 2.08911 for the Gaussian width. The position is taken to be a `visit' parameter (i.e., it is calculated once for each set of orbits corresponding to a visit), and the amplitude is taken to be an orbit parameter and is kept as a free parameter for each observation. 

In \autoref{fig:airglow variation} we illustrate the periodicity observed in both the amplitude and Doppler shift of the airglow signal. In both cases we detect a 365.25 day periodic signal (each derived independently from one another) which allows us to be confident when fitting and removing the signal from the image in order to extract the signal due to TRAPPIST-1 alone. 

\subsection{Background Model}
The background flux is modeled using a linear model on the spatial axis (the y-axis in \autoref{fig: transit fit example}), which accounts for a ramp effect across the image. The offset and slope of the ramp are found to change between each orbit, and are thus fit independently for each orbit. When analyzing the detector image, a 50x50 pixel 'stamp' is selected, centered on the Ly-$\alpha$ signal. Within this small stamp, we found that a linear ramp was sufficient to capture any background effects, and more higher-order background fitting did not improve the SNR of the retrieved values. 



\subsection{Signal Model}
The final components to describe are the emission signal coming from TRAPPIST-1, as well the attenuation due to absorption by the interstellar medium (ISM) along the line-of-sight. The Ly-$\alpha$ emission is modeled with a Voigt profile along the wavelength axis (similarly to Earth's airglow) and with a Gaussian profile in the spatial axis. The absorption due to the interstellar medium is modeled as the negative exponential of a gaussian along the wavelength axis.

\newpage

\section{Transit Lightcurves of all Planets}
\label{APP: lightcurves}

\begin{figure*}[ht]
	\centering
	\includegraphics[width=6.75in,angle=0]{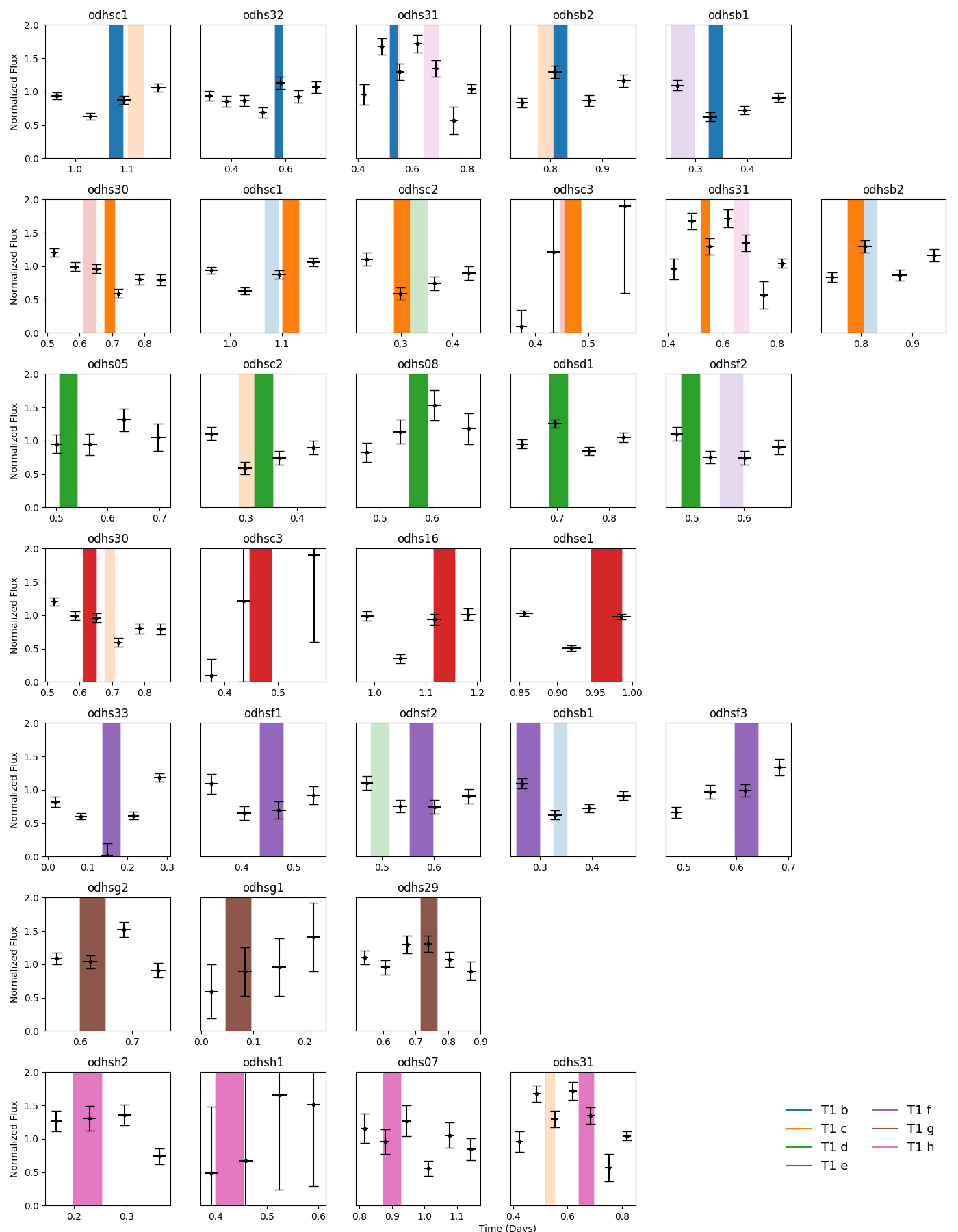}
	\caption[Ly-$\alpha$ emissions during transits of TRAPPIST-1 h]{Ly-$\alpha$ time series normalized at the individual-visit level (mean visit flux reported in \autoref{tab: STIS visits}). From top to bottom, each row shows all visits which contain transits of planets b through h (shown in in solid colors). Other planets transiting during the same visit are shown in transparent bands. The bottom right diagram shows the planet to color mapping.}
	\label{fig:all_transits}
    \vspace{-50mm}
\end{figure*}

\newpage
\section{Summary table for transit depth and scatter per planet}

\begin{table}[ht!]
\caption{Ly-$\alpha$ transit depths of TRAPPIST-1 planets} \label{tab: ly a transit depths}
\vspace{3pt}
\centering
\begin{tabular}{c c c c c}
\hline
\hline
Planet & Visits & Depth & Depth & OOT Baseline \\
 &  & (\%) & Uncertainty (\%) & Scatter(\%) \\
\hline 

T1 b & 5 & -9.5 & 10.5 & 9.68 \\
T1 c & 6 & 11.4 & 19.1 & 15.81 \\
T1 d & 5 & -13.3 & 8.5 & 5.84 \\
T1 e & 4 & -2.4 & 28.8 & 24.22 \\
T1 f & 5 & 18.9 & 14.6 & 13.54 \\
T1 g & 3 & -0.4 & 19.4 & 14.13 \\
T1 h & 3 & -9.2 & 19.3 & 17.07 \\
\hline \\
\end{tabular}
\end{table}

\section{Annual Trends from HST Motion}

\begin{figure*}[ht!]
\centering
	\includegraphics[width=7in,angle=0]{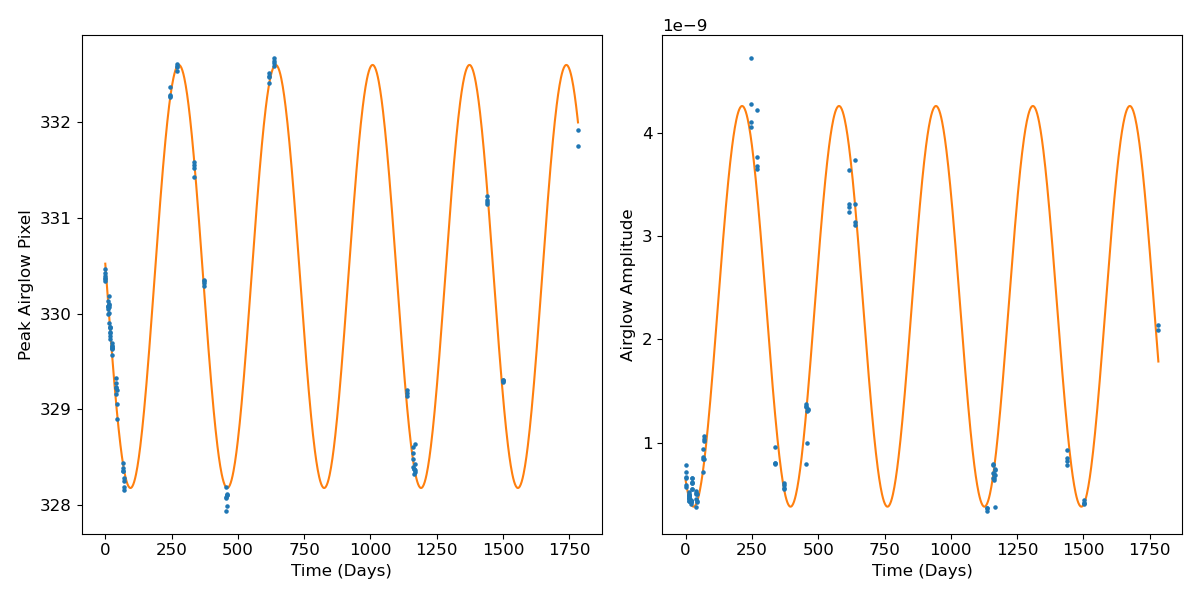}
	\caption[airglow var]{Left: The central wavelength of the peak of the airglow signal (in pixels) as a function of time. Right: The variability of the airglow amplitude as a function of time. The orange curves in both figures represent best fit sin curves, which were found to have periods of 365.25 days each.}
 \label{fig:airglow variation}
\end{figure*}

\end{document}